# Coherence analysis of local randomness and nonlocal correlation through polarization-basis projections of entangled photon pairs

Byoung S. Ham
School of Electrical Engineering and Computer Science, Gwangju Institute of Science and Technology, Gwangju 61005, South Korea
(Sep. 28, 2024; bham@gist.ac.kr)

**Abstract**
Polarization-entangled photon pairs generated from second-order nonlinear optical media have been extensively studied for both fundamental research and potential applications of quantum information. In spontaneous parametric down-conversion (SPDC), quantum entanglement between paired photons, often regarded as 'mysterious,' has been demonstrated for local randomness and nonlocal correlation through polarization-basis projections using linear optics (Phys. Rev. A **60**, R773 (1999)). This paper presents a coherence analysis of these established quantum phenomena with polarization control of the paired photons and their projection measurements. First, we analyze the quantum superposition of photon pairs generated randomly from cross-sandwiched nonlinear media, focusing on local randomness, which depends on the incoherence among measured events. Second, we investigate coincidence detection between paired photons to understand the nonlocal correlation arising from independently controlled remote parameters, resulting in an inseparable product-basis relationship. This coherence-based approach sheds light on a deterministic perspective on quantum features, emphasizing the significance of phase information intrinsic to the wave nature of photons.

**Introduction**
Quantum superposition [1,2] and quantum entanglement [2,3] are fundamental concepts in quantum information science [4]. In an interferometer, the quantum superposition of a single photon can be expressed by orthonormal bases, resulting in either the wave or particle nature depending on measurements [5-12]. As is well known, the particle (wave) nature corresponds to the distinguishable (indistinguishable) characteristics of photons, observed without (with) interference fringes. As the number of interacting photons increases, the intensity order of interference fringes can also increase, leading to multi-order intensity correlations [13-21]. In a two-photon bipartite system, these interactions can produce four distinct Bell states, determined by the relative phase between the product bases of the interacting photons [15]. However, due to the uncertainty principle, the phase of a single photon remains undetermined [3]. This does not imply that there is no phase information in a single photon; rather, the phase is random. Thus, a fixed relative phase between interacting photons can exist in a bipartite quantum system without violating the uncertainty principle, as theoretically [19] and experimentally [7,21] demonstrated. Typically, a zero relative phase is arbitrarily assigned to quantum operators [3]. For instance, the phase difference between reflective and transmitted photons on a beam splitter (BS) is often represented as $\pi$, which contradicts the basic coherence analysis of the BS for $\pi/2$ [22]. This quantum approach to single photons, which neglects phase information, can obscure the intricate quantum features related to second-order intensity correlations, such as the Hong-Ou-Mandel effect [13], Franson-type nonlocal correlations [14], and violations of Bell inequality [15].

    The wave-particle duality of a single photon is fundamental to quantum mechanics in the microscopic regime [23], where these mutually exclusive properties are determined by measurements [5-12]. Emphasizing the particle nature often leads to overlooking phase information, even in two-photon interacting bipartite systems [13-18]. The light source has a specific bandwidth defined by either linear (first order) [24] or nonlinear (second or third order) optics [25], where generated photons are phase coherent [7,12] or incoherent [13-15,21,26], depending on the generation or measurement approach [7]. The coherence property is not only individual but also effective from the photon ensemble within the frequency bandwidth [13-18,21]. Although the pump photon's phase is random, each entangled photon pair generated via spontaneous parametric down-conversion (SPDC) in a second-order nonlinear medium [25-27] is inherently phase coherent due to the phase-matching condition [21,25,27]. However, this does not mean that effective coherence among all pairs must be

preserved for the first-order intensity correlation [26]. In this paper, we analytically investigate the observed quantum features of local randomness and nonlocal correlation as discussed in ref. 26, focusing on polarization-basis control and projection measurements from the perspective of the wave nature. The derived coherence solutions aim to deepen our understanding of the 'mysterious' quantum phenomena observed.

**Methods**

Figure 1 illustrates the schematic for controlling the coherence of the polarization bases of entangled photon pairs generated from a pair of cross-sandwiched type I BBO crystals [26] through the SPDC process [25,27]. The resulting polarizations of the generated photon pairs from different crystals are orthogonal, with each pair being parallelly polarized, as indicated by the blue and red circles [26]. Positioned oppositely within the degenerate cross-section of the SPDC light cones, each photon stream (path) contains orthogonally polarized photons at equal probability amplitudes. To achieve polarization-basis control, a set of linear optics consisting of a half-wave plate (HWP) and a polarizing beam splitter (PBS) is employed for the quantum eraser-like polarization-basis projection measurements (see the green-dotted box) [7,9,21,26]. For the two-photon correlation, the two transmitted photon streams from the PBSs are coincidently measured by a set of single-photon detectors D1 and D2 [26]. Additionally, a quarter-wave plate is used to control the phase of the $\hat{V}$-polarized photons [24], allowing us to preset a specific Bell state [15,26].

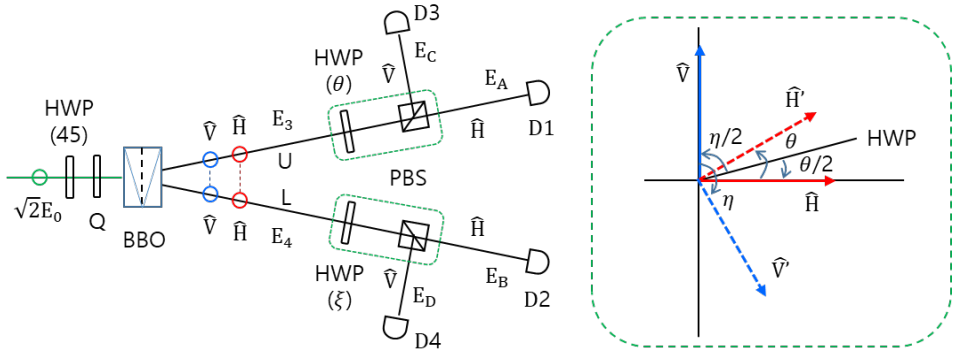

**Fig. 1**. Schematic of polarization-entangled photon pair generation. H (45): half-wave plate, HWP: half-wave plate. Inset: HWP-induced polarization-basis rotation. $\eta = \pi - \theta$.

The Inset of Fig. 1 provides details on the polarization-basis projection achieved through the linear optics setup of the HWP and PBS. This setup enables selective measurements that convert the originally orthogonally polarized photons in each path into indistinguishable characteristics (see Analysis). The HWP's role is to create a new set of orthogonal polarization bases, represented by the colored dashed arrows $\hat{H}'$ and $\hat{V}'$. These new bases are projected onto the horizontal or vertical axis of the PBS, allowing for filtering into the same polarization basis across both transmitted ($E_A$; $E_B$) and reflected ($E_C$; $E_D$) ports. For coherence analysis, the two sets of output photons from the PBSs are measured both individually and jointly to respectively access first-order and second-order intensity correlations [26]. In the case of second-order intensity correlation, each output photon pair from both PBSs is coincidently detected using a set of single-photon detectors. For first-order intensity correlation, all photons are measured individually across all detectors. Both types of measurements are facilitated by a coincidence counting module (CCU) connected to the ideally four single-photon detectors D1-D4.

**Analysis**

For the coherence analysis of polarization-basis control and projection measurements in Fig. 1, we adopt a pure coherence approach for a single photon, with its corresponding electric field represented by a harmonic oscillator governed by Maxwell's equations [24]. According to the Born rule, which states that measurements correspond to the absolute square of the amplitude, any frequency-dependent phase error of a single photon does not affect the measurement outcomes. Furthermore, the bandwidth-distributed photon pairs are generated individually and independently, forming a statistical ensemble in which the phase of the corresponding pump photons are random in the entanglement generation process. To satisfy the energy conservation law, the amplitude of the pump photon is defined as $\sqrt{2}E_0$, where $E_0$ represents the amplitude of a single photon produced by the SPDC process in a frequency-degenerate scheme. The polarization of photons in each path in Fig. 1 is random due to the configuration of the cross-sandwiched BBO crystals, with the 45 HWP introducing random polarization bases ($\hat{H}$ and $\hat{V}$) for the pump photon [26]. Consequently, the diagonally polarized pump photons indicated by the green circle are converted into orthogonally polarized photon pairs with equal probability amplitudes (see the blue and red circles). The polarization direction of the paired photons is either horizontal or vertical, determined by the pump photon's polarization [26]. Thus, each path-polarization entangled photon pair generated from the cross-sandwiched BBO crystals can be expressed by $|E\rangle_{en} = E_0\left(|\hat{H}\hat{H}\rangle_{UL} + e^{i\psi}|\hat{V}\hat{V}\rangle_{UL}\right)/\sqrt{2}$, where $\psi$ is deterministically controlled by the tilting angle of the QWP (Q) [26].

For $\psi = \pi$, the following amplitude-superposition relation is coherently derived from the randomly and independently generated photons in each path of the SPDC light cones:

$$\vec{E}_3 = \vec{E}_4 = E_0(\hat{H} - \hat{V})/\sqrt{2}, \qquad (1)$$

where the unit vectors for horizontal and vertical polarizations are denoted by $\hat{H}$ and $\hat{V}$, respectively. The corresponding intensities of Eq. (1) are $\langle I_3\rangle = \langle I_4\rangle = I_0$, where $I_0 = E_0 E_0^*$, satisfying the distinguishable characteristics of the photon ensemble [26]. However, the source of this particle (distinguishable) nature remains unclear, whether it stems from the orthogonal polarization bases or incoherence among them. Therefore, it is essential to investigate the randomness of the first-order intensity correlation to determine its origin. This constitutes the first goal of this paper.

Each photon pair generated via SPDC from the BBO crystals in Fig. 1 is now subjected to polarization-basis control using the HWP and PBS in both paths. The HWP rotates the polarization axis of an incident single photon, as indicated by the dashed-color arrows in the Inset of Fig. 1 [24]. The role of the PBS in performing polarization-basis projection through the HWP is akin to the quantum eraser function of a polarizer (see Supplementary Materials) [9,12], where orthogonally polarized photons are projected onto a common polarization basis of the PBS. Consequently, the combination of linear optics with the HWP and PBS yields indistinguishable photons for measurements. Thus, the coherent derivation of the polarization-basis projection onto $\hat{H}$ of the PBS is expressed as follows:

$$\vec{E}_A = E_0(Hcos\theta - Vsin\theta)\hat{H}/\sqrt{2}, \qquad (2)$$

$$\vec{E}_B = E_0(Hcos\xi - Vsin\xi)\hat{H}/\sqrt{2}, \qquad (3)$$

where $\theta/2$ ($\xi/2$) represents the rotation angle of the HWP in the upper (lower) path (see the Inset) [24]. The dummy notations with H and V in Eqs. (2) and (3) indicate the origin of the photon's polarization for subsequent analysis; they no longer imply "orthogonal."

The corresponding intensities of Eqs. (2) and (3) are as follows:

$$I_A = I_0(1 - HVsin2\theta)/2, \qquad (4)$$

$$I_B = I_0(1 - HVsin2\xi)/2, \qquad (5)$$

where the product basis $HV$ is supposed to be permitted through the polarization-basis projection onto the PBS [9,12]. As in the quantum eraser, thus, $\theta$- ($\xi$-) dependent interference fringes are anticipated [9,12]. However, the experimental observations for Eqs. (4) and (5) reveal no fringes, leading to $\langle I_A\rangle = \langle I_B\rangle = I_0/2$ [26]. Thus, this suggests that no phase coherence exists between entangled photon pairs generated from different BBO crystals in Eq. (1), resulting in $\langle \vec{E}_3 \rangle = \langle \vec{E}_4 \rangle = \frac{E_0}{N\sqrt{2}} \sum_{j=1}^{N} (\hat{H} - \hat{V}e^{i\eta_j})$ and $\hat{H}\hat{V} \sum_{j=1}^{N} cos\eta_j = 0$: For simplicity, the Gaussian distributed photons are treated as uniformly distributed. This finding highlights the first significant aspect of the proposed coherence analysis, where the origin of local randomness observed in ref. 26 arises from the incoherent photon characteristics of entangled photon pairs. Such polarization-basis-independent local randomness has also been experimentally observed in a Sagnac interferometer with a bidirectional pumping scheme [21] and in a double-slit BBO crystal with a unidirectional pumping scheme [28] for the type II SPDC process.

Similarly, the polarization-basis projection onto the $\hat{V}$-axis of the PBS for the orthogonally polarized pump photons can be coherently derived as follows:

$$\vec{E}_C = E_0(\text{H}sin\theta + Vcos\theta)\hat{V}/\sqrt{2}, \quad (6)$$

$$\vec{E}_D = E_0(\text{H}sin\xi + Vcos\xi)\hat{V}/\sqrt{2}, \quad (7)$$

where the corresponding intensities are $\langle I_C \rangle = \langle I_D \rangle = I_0/2$ due to the same reason of the incoherent photon ensemble, as in Eqs. (4) and (5). The measurement results in ref. 26 are frequency-independent for all frequency-distributed photon pairs, as defined by the SPDC process. The only matter is the Gaussian distribution, which leads to differing probabilities. Thus, the sum of all four intensities individually measured for the statistical ensemble must comply with the energy conservation law: $\langle I_T \rangle = \langle I_A \rangle + \langle I_B \rangle + \langle I_C \rangle + \langle I_D \rangle = 2I_0$.

For the intensity product between two presumably space-time separated photons detected by D1 and D2 in Fig. 1, the coincidence detection $R_{AB}^{(H)}(0)$ is directly derived from Eqs. (2) and (3):

$$\langle R_{AB}^{(HH)}(0) \rangle = \langle (E_A E_B)(cc) \rangle = \langle (\text{H}cos\theta - Vsin\theta)(\text{H}cos\xi - Vsin\xi)(cc) \rangle I_0^2/4$$

$$= \langle (HHcos\theta cos\xi + VVsin\theta sin\xi - HVcos\theta sin\xi - VHsin\theta cos\xi)(cc) \rangle I_0^2/4$$

$$= \langle (cos(\theta - \xi) - HVsin(\theta + \xi))(cc) \rangle I_0^2/4$$

$$= I_0^2 \langle cos^2(\theta - \xi) \rangle/4, \quad (8)$$

where cc denotes the complex conjugate. For the space-time separation, the phase coherence between paired photons is always kept within the coherence length of the difference length. In Eq. (8), a fixed relative phase $e^{i\zeta}$ between all paired photons is canceled by cc. Unlike Eqs. (4) and (5), Eq. (8) is for entangled photons in each pair, satisfying the statistical ensemble in average.

Equation (8) represents the typical form of nonlocal correlation [15], satisfying the violation of Bell's inequality [29]. Similar to the local measurements of Eqs. (4) and (5), the nonlocal product-basis term $HV[sin(\theta + \xi) + sin\theta cos\xi]$ is not allowed, either, but for a different reason related to the lack of simultaneous generation. The SPDC process yields either an *H-H* or *V-V* entangled photon pair, as indicated by the colored circles in Fig. 1. Thus, the coincidence detection acts as a selective measurement process that eliminates the *HV* product basis term. As derived in Eq. (8), the analytical solution for the nonlocal correlation observed in ref. 26 is the direct result of the wave nature of entangled photons with a fixed relative phase between paired photons. The fixed phase relation between paired photons is crucial for understanding the inseparability of the product basis in Eq. (8); otherwise, the nonlocal fringes would be washed out, as shown in Eqs. (4) and (5). This is the second significance of this coherence approach.

Unlike single photons, assigning a relative phase to the paired photons does not violate quantum mechanics or the coincidence measurement (due to QWP adjustment) in Fig. 1. With this phase information, Eq. (8)

supports the wave nature interpretation of the observed two-photon correlation [26]. Thus, the observed nonlocal correlation in ref. 26 is effectively analyzed for the deterministic process of local variables $\theta$ and $\xi$ of the HWPs. Likewise, the correlation for the reflected photon pairs detected by the PBSs can be expressed as $\langle R_{CD}^{(VV)}(0)\rangle = \langle R_{AB}^{(HH)}(0)\rangle$ according to Eqs. (6) and (7) (see the Supplementary Materials).

Similarly, the intensity product between D1 and D4 can also be coherently derived for the same nonlocal correlation using Eqs. (2) and (7):

$$\langle R_{AD}^{(HV)}(0)\rangle = \langle E_A E_D(cc)\rangle = \langle(H\cos\theta - V\sin\theta)(H\sin\xi + V\cos\xi)(cc)\rangle I_0^2/4$$

$$= \langle(HH\cos\theta\sin\xi - VV\sin\theta\cos\xi + HV\cos\theta\cos\xi - VH\sin\theta\sin\xi)(cc)\rangle I_0^2/4$$

$$= \langle(\sin(\xi - \theta) + HV\sin(\theta + \xi))(cc)\rangle I_0^2/4$$

$$= \langle\sin^2(\xi - \theta)\rangle I_0^2/4. \tag{9}$$

For the other pair of transmitted and reflected photon pairs by the PBSs, $\langle R_{BC}^{(VH)}(0)\rangle = \langle R_{AD}^{(HV)}(0)\rangle$ is coherently derived from Eqs. (3) and (6). As numerically simulated in Fig. 2, Eqs. (8) and (9) exhibit an out-of-phase relationship. Thus, the sum of Eqs. (8) and (9) reveals a classical feature characterized by uniform intensity (see the right column of Fig. 2): $\langle R_{AB}^{(HH)}(0)\rangle + \langle R_{AD}^{(HV)}(0)\rangle = \langle R_{CD}^{(VV)}(0)\rangle + \langle R_{BC}^{(VH)}(0)\rangle = I_0^2/2$. This expression represents the classical intensity product of Eqs. (4) and (5) for the whole events without coincidence detection. Thus, the quantum features obtained in Eqs. (8) and (9) raise the question of "*how to measure*." Similar to the spectral filtering of sunlight through a prism to create colored lights, this "how to measure" question implies "how to select" the measurable events. In other words, the nonlocal quantum feature becomes less mysterious when viewed through the lens of coherence and a deterministic, controllable measurement process. Here coincidence detection serves as a classical method. Another relevant example of this type of measurement choice is the Franson-type nonlocal correlation [14], which has been coherently analyzed, too [19]. Thus, we now understand that measurements play a critical role in quantum mechanics, where the nonlocal quantum feature comes at the cost of a 50% loss in measurement events due to the selective choice of specific product bases.

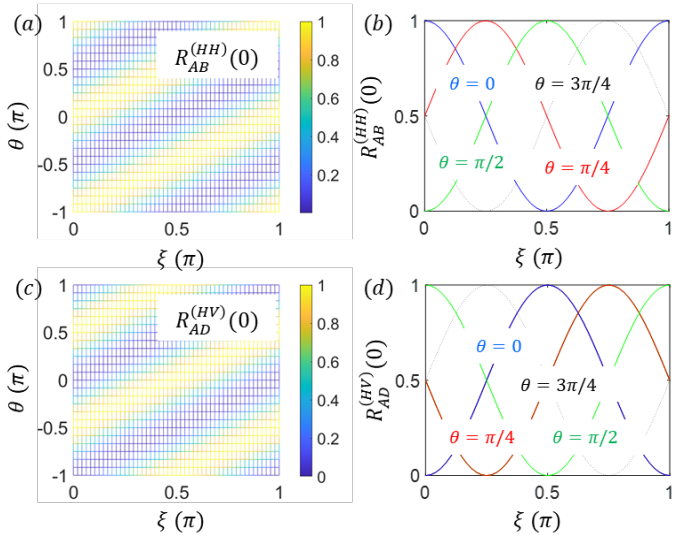

**Fig. 2**. Numerical calculations of the two-photon correlation in Eq. (8). $R_{CD}^{(VV)}(0) = R_{AB}^{(HH)}(0)$. $R_{BC}^{(VH)}(0) = R_{AD}^{(HV)}(0)$.

Figure 2 represents numerical calculations of the analytical solutions derived in Eqs. (8) ad (9). The top row of Fig. 2 is for the transmitted photon pairs in Eq. (8), confirming the inseparable product-basis relationship in the nonlocal quantum feature, which satisfies the violation of Bell's inequality [15]. The bottom row shows the out-of-phase relationship with respect to the top row, where the sum of the same colored curves yields a classical feature of uniform intensity, independent of the local parameters $\theta$ and $\xi$. Thus, it is numerically confirmed that the selective measurement acts through coincidence detection to extract the quantum feature from classical events.

**Discussion and conclusion**

In conclusion, the polarization-basis control of type I SPDC-generated photon pairs from sandwiched BBO crystals was coherently analyzed for both local randomness and nonlocal correlation using a set of linear optics comprising HWP and PBS. As derived in Eqs. (4) and (5), the origin of local randomness was attributed to the incoherence among photon pairs, consistent with the statistical ensemble in quantum mechanics. Furthermore, as derived in Eqs. (8) and (9), the polarization-basis control of the paired photons enabled selective measurement to eliminate orthogonal product basis, a behavior not typical in classical measurements. This control was achieved through linear optics, projecting the orthogonally polarized photons onto a common polarization basis, which resulted in indistinguishable photon characteristics. The fixed relative phase between entangled photons was crucial in this process to avoid the coherence washout. If such polarization-basis control and projection measurements could be realized with conventional laser light, it would also enable macroscopic quantum features, thanks to the same coherence approach. Consequently, quantum measurements would be applicable in a single shot, as demonstrated in the NV-center diamond [30] and rare-earth doped solids [31]. Thus, the coherence solution for the quantum features derived in Eqs. (8) and (9) shed light on our understanding of quantum mechanics, clarifying the origins of local randomness and nonlocal fringes.